# Applying Artificial Intelligence to Clinical Decision Support in Mental Health: What Have We Learned?

**Authors:** Grace Golden[1,2], BASc; Christina Popescu[2,3], BSc, MSc; Sonia Israel[2], BSc; Kelly Perlman[2,3], BSc; Caitrin Armstrong[2], BASc; Robert Fratila[2], BSc; Myriam Tanguay-Sela[2], BASc, & David Benrimoh[2,3,4], MD.CM., MSc., MSc., FRCPC

**Affiliations:** [1]University of Western Ontario, London, ON, Canada; [2]Aifred Health Inc., Montreal, QC, Canada; [3]McGill University, Montreal, QC, Canada; [4]Stanford University, Palo Alto, California, United States

**Author Contribution:**
GG and DB conceptualized the manuscript and contributed to the manuscript writing and review. CP, SI, KP, CA, RF, and MTS contributed to the manuscript writing and review.

**Abstract**: Clinical decision support systems (CDSS) augmented with artificial intelligence (AI) models are emerging as potentially valuable tools in healthcare. Despite their promise, the development and implementation of these systems typically encounter several barriers, hindering the potential for widespread adoption. Here we present a case study of a recently developed AI-CDSS, Aifred Health, aimed at supporting the selection and management of treatment in major depressive disorder. We consider both the principles espoused during development and testing of this AI-CDSS, as well as the practical solutions developed to facilitate implementation. We also propose recommendations to consider throughout the building, validation, training, and implementation process of an AI-CDSS. These recommendations include: identifying the key problem, selecting the type of machine learning approach based on this problem, determining the type of data required, determining the format required for a CDSS to provide clinical utility, gathering physician and patient feedback, and validating the tool across multiple settings. Finally, we explore the potential benefits of widespread adoption of these systems, while balancing these against implementation challenges such as ensuring systems do not disrupt the clinical workflow, and designing systems in a manner that engenders trust on the part of end users.

**INTRODUCTION**

Clinical Decision Support Systems (CDSS) are computerized point-of-care systems intended to support clinical decision making.[1] These systems consolidate large quantities of clinical information and are intended to improve healthcare delivery by improving clinicians' decision-making.[2,3] CDSS are emerging tools with the potential to impact both physical and mental healthcare[4], but their implementation in clinical practice remains limited. The current article explores the barriers and challenges concerning CDSS development and implementation, provides recommendations to consider throughout the process, and discusses the potential benefits of widespread adoption, balanced against the challenges faced in this process. A key theme we will develop is the need for continuity between the development and design of a CDSS and its clinical implementation.

**Types of Clinical Decision Support Systems**

CDSS have a long history and have been developed for various physical and psychiatric medical conditions (refer to Supplementary for more detail).[5,6,7] There are two types of CDSS: knowledge-based (KB) and non-knowledge-based (NKB) systems.[1,3] A KB CDSS is built on static evidence-based rules and clinical knowledge and may provide diagnostic or treatment suggestions or reminders.[1,3] Meanwhile, NKB systems incorporate some form of a statistical model that generates outputs based on observed data.[3] A subset of these systems incorporate complex statistical learning models known as artificial intelligence (AI). A NKB CDSS can also contain a KB CDSS (e.g., a CDSS that provides predictions may also include information from guidelines). Here, we will focus the discussion on NKB CDSS with integrated AI models (termed AI-CDSS), which also have elements of KB systems.

# AI-CDSS: BARRIERS AND CHALLENGES

*Integration of AI into CDSS*

Research on the application of AI for medical predictions is increasingly popular.[8,9,10,11] To date, most of these AI models have been developed as standalone tools not integrated into a CDSS, which may reduce adoption by the broader clinical community. One possible reason is that most models are generated because of basic research efforts in which model construct is the primary goal, rather than within a translational research context, which prioritizes the generation of a clinically useful tool.[12] Models developed in a basic research context may not transition to a CDSS for several reasons. For instance, there may be difficulties with the model itself (e.g., creating a model that does not generalize to a target population), or a failure to integrate the model into the clinical workflow (e.g., a model which requires complex tests as inputs which are not routinely collected in clinical practice).[13]

*Considerations for Selecting Data Sources*

Successful integration of an AI model into CDSS requires that the model reach a field-specific standard of accuracy based on varying levels of acceptable risk and uncertainty in different domains. Generating accurate models depends on having sufficient data of a high enough quality, which addresses the intended population at time points relevant to the decision that needs to be made. Quality begins with obtaining large amounts of data, often from electronic medical records (EMR) or clinical trial repositories, as small datasets have problems with biased outputs or difficulties with generalizability.[3,11] Even if a model is accurate and reliable when predicting a given endpoint, choosing that endpoint is essential as it will determine whether the model will have clinical utility. For example, imagine a model that accurately predicts a treatment outcome in a disease with only one effective treatment. In this case, knowing the predicted outcome likely will not change the clinical management as clinicians are unlikely to deny their patients a chance to try the treatment. In addition, a model that makes predictions between multiple treatments may have greater utility if there is no clinical reason to select one treatment over another than if guidelines or clinical experience dictates a specific order in which to try treatments.

*Adoption into Clinical Practice*

When discussing the success or failure of CDSS, the system needs to be judged by metrics specific to the field and, indeed, to the decision(s) they are intended to assist with. One potentially universal metric is adoption (i.e., the number of clinicians who regularly use the CDSS); while high adoption does not guarantee impact, it is a prerequisite for impact at scale. The lack of CDSS in usual clinical practice, and especially the lack of AI-CDSS, is striking, as it indicates that even if useful CDSS exist, they do not currently have a significant clinical impact because of low adoption. Why might this be? CDSS have been criticized for taking too much time[14], have been received by clinicians with uncertainty due to low confidence in the AI-produced results and a lack of overall trust in the system[15,16,17], there has been resistance within the medical field to accept AI, stemming, at least in part, from the lack of transparency concerning machine learning (ML) models.[18,19] Given the nature of medicine, which partially relies on mechanistic information, physicians may not trust the reasoning and decisions behind the model prediction if the system does not provide justifications for why or how a prediction was determined.[18,19] Lack of trust is the "black box" problem of AI, and research on AI interpretability methods is now a primary focus of the field.[20]

An important criticism is that incorporating CDSS into clinical practice disrupts or interferes with workflow and often results in physician frustration and burnout, possibly due to a lack of training or support.[14,21] These examples demonstrate how despite the potential of CDSS to improve clinical care and patient outcomes, improper design or implementation can introduce new errors or problems, decrease tool retention, or prevent adoption entirely.[22]

*Identifying a Relevant Problem*

Finally, it is important to consider a fundamental premise of CDSS design: CDSS should not be developed for the sake of adopting new technology but rather should address what clinicians need assistance with, focusing on solving current problems that do not already have simpler or more effective solutions.[18]

The focus must be on supporting the decisions for which physicians require assistance, which is a useful starting point for the case study we will focus on for the remainder of this paper. Consider treatment selection and management in major depressive disorder (MDD). While there are several medications, psychotherapies, and neurostimulation techniques available to treat depression[23], it remains the case that two-thirds of patients do not achieve remission after their first course of treatment.[24] Patients and their physicians often face a trial-and-error approach to treatment selection.[25] Furthermore, physicians usually become accustomed to prescribing the same limited treatments (potentially because of their familiarity with side effects and typical outcomes), which results in a few treatments, often medications, being used as the initial treatment for patients.[25,26] Beyond treatment *selection*, there are challenges in treatment *management*: many clinicians do not routinely follow guidelines, despite research demonstrating that when algorithms are used, patients experience improved outcomes.[27,28] As such, in the case of MDD, there seems to be room for (1) a KB CDSS to address treatment management and (2) a NKB CDSS to address treatment selection. For a CDSS to solve this problem, it must be developed to assist clinicians in personalizing treatment choices and then assisting the clinician in managing the treatment.[18,25] Treatment selection and management is the challenge we set ourselves when creating the Aifred Health CDSS.

**BUILDING A CDSS**

**Basic Science and Design**

The first step in building our CDSS was considering what features clinicians and patients required via discussions with clinician and patient stakeholders and a survey of the evidence-based depression treatment literature. We decided to solve treatment selection using AI, as existing approaches that used classical statistics had failed to generate reliable tools to help personalize treatment.[25,29,30] We determined that improving treatment management did not require an AI approach, as existing guidelines provided clear recommendations and the use of measurement-based, algorithm-guided care has been proven to improve patient outcomes but simply is not being implemented with sufficient fidelity in clinical practice (see Table 1). [23,31]

The design of the AI component took place concurrently with the design of the patient and clinician user experience (see Figure 1). The resulting application was influenced by what clinicians and patients explained they needed, and the kinds of outputs generated by the AI were engineered to fit into the clinical workflow and to be both useful and more easily interpretable to a clinician. Furthermore, to prevent clinician and patient burden and reduce patient and clinician discontinuation rates, the AI required as few inputs as possible.[32,33]

Furthermore, it was clear from early discussions that clinicians did not simply want a tool to give an overall patient probability of remission; rather, they wanted one to help them select between treatments. Clinician input is why a focus of the model was to allow for differential treatment benefit prediction - that is, models that allow for comparison between multiple individual treatments.[30] Based on clinician feedback, in later iterations of the model, we focused on creating a model that generated clinically meaningful differences in predicted efficacy between the best and worst treatments for each patient.[34]

**Figure 1**

*Clinical Decision Support System Development, Implementation, and Validation Process*

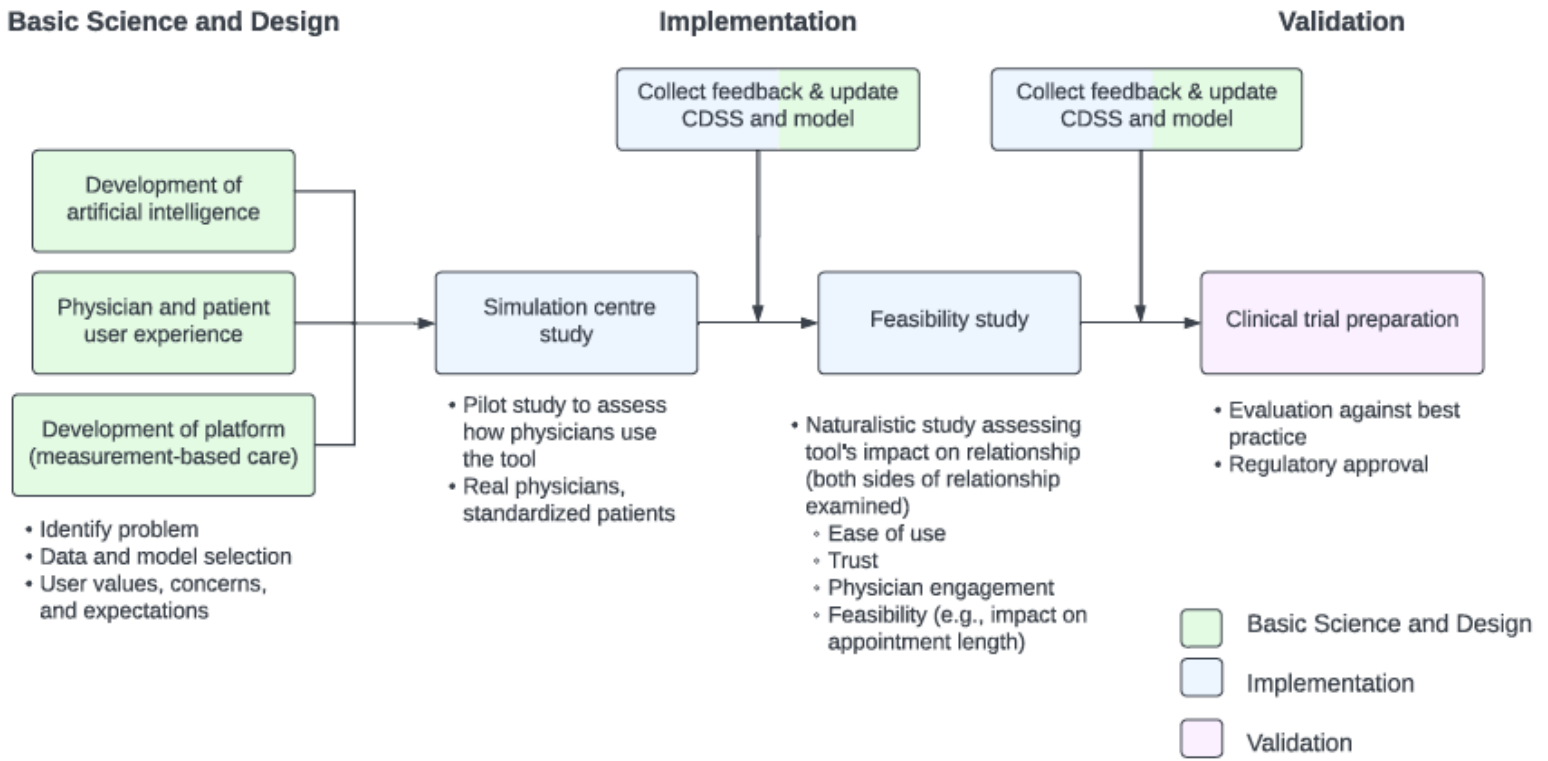

## Integration of AI into CDSS

Once we understood what problem the AI needed to solve, we also needed to determine how to incorporate it into the system. To assist with treatment management, the CDSS contains an operationalized version of the Canadian Network for Mood and Anxiety Treatments (CANMAT) 2016 guidelines.[23] These operationalized guidelines provide information based on the patient's current status and include a treatment selection module where clinicians are given information about treatments listed in the guidelines. At this step, we presented the AI predictions *within* the guideline module but visually differentiated the AI from the guideline-derived information. To minimize the impact of the system on physician autonomy, the AI generates a probability of remission for each treatment in an ordered list rather than providing a single recommendation, prompting the clinician to discuss treatment options with the patient to encourage shared decision making.[34]

As noted above, a key barrier to physician adoption of AI tools is a lack of trust in these systems because of a lack of interpretability. To address this, we created the "interpretability report" (described in Mehltretter[30]), which generates the five patient variables most strongly associated with the predicted probability of remission for each drug. Physician-perceived relevance of this report to a simulated patient was correlated with physician trust, which interacted with patient severity to predict physician prescription of one of the top treatments predicted by the AI in a simulation center study.[34,35]

As we designed the CDSS that would house the AI, we simultaneously developed the AI itself. We began with a comprehensive review of which predictors could influence treatment outcomes in patients with MDD to determine what our model should include and to create a reference against which associations learned by the model could be compared. Initial predictors identified included symptoms, demographic information, and various biomarkers.[36] However, after reviewing available data, we determined that there is a paucity of biomarker data for many individual treatments commonly used in practice and that it was unlikely that physicians would routinely collect this data from patients using the tool.[36] As such, we focused model training solely on easy-to-collect clinical and demographic characteristics. Part of the development process involved determining which ML method to use. We ultimately chose deep learning because it can find complex, non-linear patterns in data that classical statistical models struggle to find.[37] In addition, as new predictor modalities become more feasible, integration into the core model will be easier (see Mehltretter[29] for more information).

***Considerations for Selecting Data Sources***

An important decision when creating our model was the choice of the data source. Data sources must cover the range of relevant predictors, include a variety of patients treated with different medications, and clinical characteristics (e.g., comorbidities) which approximate a real-world population. When considering clinical trial data, biases may be found due to strict inclusion and exclusion criteria (e.g., participants are not representative of patients seen in the clinic). Alternatively, while EMR systems have a surplus of available data, medical records may lack rigorous outcomes measures[38], data may be incomplete, missing, unstructured, or contain errors[39], or data may engender false inferences due to misinterpretation of consecutive visits or irregularity of visits.[40] EMR data can also be affected by temporal biases, wherein as time goes on and treatment standards change, patients from different periods may not be comparable.[41] There is also the risk of propensity biases, where clinicians may be more likely to treat certain patients in a certain way.[42] Importantly, strategies such as propensity models, can help overcome such issues and should be considered when using EMR data.[42]

Ultimately, we determined that the biases present in clinical trial data were easier to identify and investigate in practice than EMR data and that sufficient data was available from trials with looser inclusion/exclusion criteria that a dataset approximating a real clinical population could be created. In addition, a clear advantage for clinical trial data was the presence of unambiguous outcomes which could be used as training targets.[29,30,35]

**Implementation**

*Adoption into Clinical Practice*

Implementing the tool into clinical practice requires validating the tool in a clinical setting, which is necessary to build trust between the physician and the AI-powered tool.[25] Pilot and quality assurance testing also provide the opportunity to find and address software errors or mismatches between the tool's design and the clinical workflow before larger trials, thereby hopefully reducing the incidence of adverse events and reducing the impact of suboptimal design on the performance of the CDSS in a clinical trial.

Given the importance of adequate and iterative testing during the implementation phase, one of our earliest studies used a simulation center to receive feedback from physicians on their experience concerning the product's utility before incorporating it into actual clinical practice (see Benrimoh[35] and Tanguay-Sela[43] for more information). We found that clinicians noted the tool could be helpful for shared decision making with patients.[35] Importantly, given those early conversations with clinicians focused on concerns about the time using the tool would take, it was found that the tool could be used successfully during a 10-minute session with a standardized patient, which suggested that the tool was ready for feasibility testing in the clinic.

Our next step was a feasibility study of the CDSS in a real clinical environment before a clinical trial focused on effectiveness. The main focus of this study was time spent in each appointment, as clinician stakeholders mentioned that lost time would be a significant barrier to them using the CDSS in practice (see Popescu[34] for more information). We found no difference in appointment length before and after the introduction of the tool. The study also demonstrated that physicians and patients sustained engagement with the tool beyond two weeks (a two-week benchmark was used to signify retention success based on results in Arean et al., 2016)[44], potentially because the app was tied into clinical care.[34] Crucially, we found that the tool did not negatively impact the clinician-patient relationship for any of the patients – and in roughly half of all cases (i.e., 46%), the use of the CDSS was even reported to improve the relationship.[34] Such findings were important as they support that adding a "third party" (the AI-CDSS) to the clinician-patient relationship will not negatively impact it and in turn, increase the feasibility of the tool and the efficacy of treatment overall.[45]

**Clinical Trial Preparation**

The next step in the clinical validation of the tool is conducting a randomized control trial (RCT) to assess the tool's effectiveness and safety. Secondary endpoints collected will also allow continued feasibility assessment and impact on care processes. This is an important step since many CDSS are not evaluated for impact on patient outcomes in RCTs.[12]

This ongoing trial (NCT04655924) is a cluster (physician)-randomized, patient and rater-blinded, active-controlled trial of our CDSS. We designed the study to be as realistic and pragmatic as possible, meaning that physicians are free to use the AI suggestions or disregard the information it provides. Similarly, physicians receive training on best practices and are provided with questionnaire results but are not required to follow guidelines strictly. This pragmatic design was chosen to generate results relevant to real-world implementation, as it is not reasonable to assume that clinicians will use the CDSS according to a pre-defined research protocol once the study is over.

DISCUSSION

**Potential Benefits of CDSS Adoption**

CDSS have the potential to improve healthcare efficiency and reduce long-term costs.[46] A tool that augments the healthcare providers' decision-making with evidence-based support could help address the need for mental health services, potentially empowering more first-line practitioners, such as nurse practitioners and family physicians, to provide high-quality care. In addition, an AI-powered CDSS that could reduce the number of treatments that need to be tried by increasing personalization while improving the quality of treatment management has the potential to reduce the time patients remain ill, which would have a significant effect on patient and family suffering in addition to an important reduction in the strain on struggling healthcare systems and economies.[25]

**Addressing AI-CDSS Limitations**

Despite the potential benefits of implementing CDSS, multiple CDSS have often failed during the implementation stage due to a lack of transparency, adding time to routine practice, uncertainty relating to the evidence and lack of trust in the system, or disruptions to the clinical workflow.[14,15,16,17,18] The seeds of these setbacks are often sown during the initial conceptualization stage if the tool is developed to create something innovative but unsuitable for current needs.[18] We argue that the first step towards avoiding these pitfalls is to determine the specific problem to be addressed and to use this to drive the system's design - both in terms of user experience and the design of the AI model. In addition, we argue that a successful CDSS will undergo a rigorous testing and validation process.[47] In addition, this process must be iterative and include, at every stage, opportunities for incorporating user feedback. It may be tempting to begin with a large clinical trial since it will reduce the time from the initial system development to clinical implementation. However, clinical trials are expensive, resource

intensive- and expose large numbers of patients- and, therefore, should only be conducted when there is confidence in probable success and lack of harm.

**CONCLUSION**

This paper discusses current barriers which limit the widespread adoption of CDSS into healthcare and provides recommendations to consider throughout the building, validation, training, and implementation process. CDSS are innovative and efficient tools with the potential to improve healthcare delivery at the patient and system levels but have significant barriers to implementation that must be addressed to be successfully implemented.

Table 1. Unique Challenges of Building an AI CDSS and How Aifred Approached the Problem.

| Challenges | Aifred's Approach |
|---|---|
| Ensuring the CDSS solves the *right* problem.[18] | Identified treatment selection and management as key problems; treatment selection required the development of an AI model, while management required the digitization and operationalization of guidelines. |
| Data needs to be of high quality.[3,38,39,40] | Did not include EMR data due to inconsistencies in quality and availability of outcomes. Trained and validated model using baseline clinical and demographic data from antidepressant clinical trials. |
| Must validate a ML model.[29] | Developed different ML model types and compared with existing published models. Used metrics which are familiar to clinicians to increase trust. |
| Functionality must have clinical utility.[18,25] | Given that there are various medication options for MDD and clinicians struggle to choose between them, we developed an AI-powered CDSS which implements differential treatment benefit prediction to assist with selection between treatments. The CDSS also incorporates an operationalized version of the CANMAT guidelines to assist with treatment management by integrating it more effectively into the clinical workflow than is possible in standard care. |
| Developing physician trust in AI.[15,16,17,18] | Developed the interpretability report, which provides up to five patient variables that were most significant in determining the probability of remission for a particular drug; conducted stakeholder needs assessments to understand what physicians wanted from the tool and how they would use it; underwent a rigorous clinical validation process and reported results. . |
| CDSS cannot interfere in the relationship e.g., high-jacking decision-making process.[18,25,34] | Investigated tool's impact on the physician-patient relationship in simulated clinical interactions between physicians and standardized patients and during a clinical feasibility study. |

**Funding:** This research was funded by the Aifred Health Inc; Innovation Research Assistance Program, National Research Council, Canada; ERA-Permed Vision 2020 supporting IMADAPT; Government of Québec Nova Science; and the MEDTEQ COVID-Relief Grant.

**Conflicts of Interest:** GG and CP have been or are employed or financially compensated by Aifred Health. MTS is employed by Aifred Health and is an options holder. SI, KP, CA, RF, and DB are shareholders and employees, directors, or founders of Aifred Health.

**Data Availability Statement:** There is no relevant data as this is a perspective piece.

Supplemental Material

**Introduction**

*Historical Context*

First, we will provide some context with respect to the development and use of CDSS in healthcare. The United States National Health Care Act endorses the development of CDSS given their cost-effectiveness and ability to integrate into some electronic medical record (EMR) systems (Sutton et al., 2020). CDSS have various functions and advantages such as their ability to reduce prescription errors, mitigate adverse events, reduce costs (e.g., reduce test and order duplication), provide diagnostic and decision support, improve clinical workflow, deliver reminders and alerts, and improve patient outcomes (Khairat et al., 2018; Sutton et al., 2020).

Since the 1970s (Shortliffe & Buchanan, 1975), CDSS have been developed for a range of physical and psychiatric medical conditions (Jia et al., 2016; Kwan et al., 2020; Roshanov et al., 2011). Notwithstanding increasingly sophisticated clinical or data-driven algorithms and models, increased data volumes from studies and EMRs, and improvements in computational storage and power, CDSS effectiveness varies considerably with each tool (Kwan et al., 2020; Roshanov et al., 2011). For instance, a systematic review and meta-analysis conducted by Kwan et al. (2020) reported that of 108 studies, CDSS resulted in 5.8% more patients receiving care adherence to the standards programmed into the CDSS. However, the authors noted substantial heterogeneity among the top quartile of studies, ranging from 10% to 62% on improved process adherence, which suggests that CDSS interventions can have a wide range of possible impacts. Another systematic review by Roshanov et al. (2011) investigated whether CDSS improved chronic disease management processes and associated patient outcomes, and found that systems addressing diabetes and dyslipidemia showed potential for improving patient outcomes. Nevertheless, systems addressing hypertension, asthma, chronic obstructive pulmonary disease, cardiac conditions, and other care rarely demonstrated benefits with respect to patient outcomes, though studies in these conditions suggested that CDSS may improve the quality of care provided to patients (Roshanov et al., 2011). Improvements of care processes may result from the CDSS' push features (e.g., a notification message that pops up on a mobile device), complex decision support (e.g., guideline-based feedback and recommendations for screening, diagnosis, and prescribing), or documentation options (Kwan et al., 2020). This observation was corroborated by another review assessing the effects of CDSS on medication safety, where 75% of trials demonstrated positive impacts on the process of care, but only 20% demonstrated positive impacts on patient outcomes (Jia et al., 2016). Roshanov et al. (2011) noted that only 56% of trials investigated patient outcomes as the primary endpoint, indicating that CDSS, or the studies aimed at investigating their impact, are not necessarily designed with the intent to improve patient outcomes. Rather, improving the care process is often *assumed* to improve outcomes or other important metrics. For instance, a CDSS may be built with the intent to allow clinicians to see more patients by improving their efficiency or capacity, with the expectation that this will improve outcomes overall, as more patients receive care. Alternatively, study designers may be uncertain regarding the expected effect of CDSS on patient outcomes and therefore focus on assessing the systems' feasibility as the primary endpoint, without further studying the effects on outcomes. This distinction is relevant because there are several potential barriers that may prevent the translation from improved care processes to improved patient outcomes. These might include a lack of confidence by end-users in the appropriateness of outcomes being measured or of the decision support being offered, insufficient time for training or use of the CDSS during regular practice, or limited CDSS features that do not meet clinical needs. Additionally, many CDSS are built with patient-facing components and therefore, it is critical to consider the patient's perception of the CDSS during the design process (Kawamoto et al., 2005), as patient confidence in the system or their perception of its utility may drive their adoption and, in turn, the overall effectiveness of the system.

**Implementation**

*Adoption into Clinical Practice*

Given the importance of adequate and iterative testing during the implementation phase, one of our earliest studies used a simulation center to receive feedback from physicians on their experience concerning the product's utility prior to incorporating it into real clinical practice (see

Benrimoh et al., 2021 and Tanguay-Sela et al., 2022 for more information). Aifred's simulation center study assessed the CDSS among 20 participants, 11 of which were psychiatrists and 9 of which were family physicians. The study allowed us to understand how physicians would use the system in session with patients by enabling us to observe them using it with three standardized patients (one with mild, one with moderate and one with severe depression). Physicians were given basic training to use the tool, and were instructed to use the tool however they thought was appropriate during the session (Benrimoh et al., 2021). The simulation provided several insights concerning how physicians received the tool. Key among these was that clinicians noted that the tool could be helpful in elements of shared decision making with patients and that the standardized patients were most appreciative of the tool when they were 'invited in' by a clinician sharing their screen of the app with them (Benrimoh et al., 2021). 50% of physicians reported that they would use the tool for all patients with MDD, and 85% of physicians thought they would use the tool for complex or treatment-resistant patients (Tanguay-Sela et al., 2022), and it was found that the tool could be used in under 5 minutes during a session with a standardized patient, which suggested that the tool was ready to be tested for feasibility in the clinic.

This assessment of readiness prompted our feasibility study of the CDSS in a real clinical environment, a "stress test" of sorts prior to a clinical trial focused on effectiveness. Our main focus in this feasibility study was time spent in each appointment, as clinicians during our stakeholder discussions earlier in development had identified lost time as being the most significant barrier to them using the CDSS in practice. Seven clinicians treated 14 patients diagnosed with MDD (Popescu et al., 2021). Comparing the baseline appointment, in which the CDSS was not used, to subsequent visits, in which the CDSS was used, there was no significant difference in appointment length (Popescu et al., 2021). Physicians were not required to use the CDSS outside of patient appointments. These results removed the potential concern that this system would increase physician workload or time spent. In addition, 62% of patients and 71% of physicians indicated that they trusted the tool, and 92% of patients and 71% of physicians felt the tool was easy to use (Popescu et al., 2021). Finally, physicians used the application during 96% of appointments, despite being told that they were free to use it or ignore it as they saw fit (see Popescu et al., 2021 for more information). Notably, both physicians and patients demonstrated sustained engagement beyond two weeks with the tool (a two-week benchmark was used to signify retention success based on results in Arean et al. (2016) where the authors noted app usage diminished significantly after 2 weeks), potentially attributed to the fact that the app was tied into clinical care (Popescu et al., 2021). Crucially, we found that the tool did not negatively impact the clinician-patient relationship for any of the patients – and in roughly half of all cases, use of the CDSS was even reported to improve the relationship (Popescu et al., 2021). This is important as there were concerns that adding in a "third party" (the AI-CDSS) to the clinician-patient relationship might negatively impact it, and in turn reduce the feasibility of the tool and the efficacy of treatment overall (Nolan & Badger, 2005).